\documentclass[12pt,preprint]{aastex}
\newcommand{\msun}{M$_\odot$}
\newcommand{\kms}{km~s$^{-1}$}
\newcommand{\tco}{$^{13}$CO}
\newcommand{\ceo}{C$^{18}$O}
\newcommand{\tcojoz}{$\textrm{\tco~}J=1\rightarrow0$}
\newcommand{\ceojoz}{$\textrm{\ceo~}J=1\rightarrow0$}
\newcommand{\cojto}{$\textrm{CO }J=2\rightarrow1$}

\shorttitle{Outflows and Magnetic Fields in L1448 IRS 3}
\shortauthors{Kwon et al.}

\begin{document}
\title{Two Bipolar Outflows and Magnetic Fields in a Multiple Protostar System, L1448 IRS 3}
\author{Woojin Kwon, Leslie W. Looney, 
Richard M. Crutcher\altaffilmark{1}, and Jason M. Kirk\altaffilmark{2}}
\affil{Department of Astronomy, University of Illinois at Urbana-Champaign, 1002 W. Green St., Urbana, IL 61801}
\email{wkwon@astro.uiuc.edu}
\altaffiltext{1}{Additionally at National Center for Supercomputing Applications}
\altaffiltext{2}{Currently at Department of Physics and Astronomy, Cardiff University}

\begin{abstract}

We performed spectral line observations of \cojto, \tcojoz, 
and \ceojoz\ and 
polarimetric observations in the $\lambda=1.3 \textrm{ mm}$ continuum and \cojto\
toward a multiple protostar system, L1448 IRS 3,
in the Perseus molecular complex at a distance of $\sim 250 \textrm{ pc}$, 
using the BIMA array.
In the $\lambda=1.3 \textrm{ mm}$ continuum, two sources (IRS 3A and 3B) were clearly detected with 
estimated envelope masses of $0.21$ and $1.15 \textrm{ \msun}$, and one source (IRS 3C) was marginally 
detected with an upper mass limit of $0.03 \textrm{ \msun}$.
In \cojto, we revealed two outflows originating from IRS 3A and 3B.
The masses, mean number densities, momentums, and kinetic energies of outflow lobes 
were estimated.
Based on those estimates and outflow features, we concluded that 
the two outflows are interacting and that the IRS 3A outflow is nearly perpendicular 
to the line of sight.
In addition, we estimated the velocity, inclination, and opening of the IRS 3B outflow
using Bayesian statistics.
When the opening angle is $\sim 20\arcdeg$, we constrain the velocity 
to $\sim 45 \textrm{ \kms}$ and the inclination angle to $\sim 57\arcdeg$.
Linear polarization was detected in both the $\lambda=1.3 \textrm{ mm}$ continuum and \cojto.
The linear polarization in the continuum shows a magnetic field at the central source
(IRS 3B)
perpendicular to the outflow direction, and the linear polarization in the 
\cojto\ was detected in the outflow regions, parallel or perpendicular 
to the outflow direction.
Moreover, we comprehensively discuss whether the binary system of IRS 3A and 3B is
gravitationally bound, based on the velocity differences
detected in \tcojoz\ and \ceojoz\ observations and on the outflow
features.
The specific angular momentum of the system was estimated as $\sim 3\times 10^{20} \textrm{ cm}^2
\textrm{ s}^{-1}$, comparable to the values obtained from previous studies on binaries and  
molecular clouds in Taurus.

\end{abstract}


\section{Introduction}
\label{sec_intro}

L1448 is a dark cloud located approximately one degree southwest from NGC 1333 in the  
Perseus cloud complex at a distance of $\sim 250$ pc \citep[e.g.][]{enoch2006}.
Three infrared sources were observed by the Infrared Astronomical Satellite 
(IRAS) and denoted as IRS 1, IRS 2, and IRS 3 by \citet{bachiller1986}.
Due to its brightness in the IRAS bands, IRS 3 has been focused on more than the others.
Meanwhile, \citet{bachiller1990, bachiller1995} revealed a well-collimated, 
large outflow originating 
from L1448-mm, located 70$''$ southeast of L1448 IRS 3.
In fact, IRS 3 is overlapped with the blueshifted lobe of the outflow.
\citet{curiel1990} detected the L1448-mm continuum source and two sources 
separated by $7''$ in the L1448 IRS 3 region, using the Very Large Array (VLA) at $\lambda=6$ and 2 cm
wavelengths.
They named the sources L1448 C (for Center) and L1448 N(A) and N(B) (for North)
for L1448-mm and the two sources in IRS 3, respectively.
\citet{terebey1997} detected another source 20$''$ northwest 
of the two IRS 3 sources at $\lambda=2.7$ mm continuum, using the Owens Valley Radio Observatory (OVRO)
millimeter interferometer.
\citet{looney2000} resolved all three sources in IRS 3 with high resolution  
Berkeley Illinois Maryland Association (BIMA) observations at $\lambda=2.7$ mm continuum. 
In this paper, we call these sources L1448-mm and L1448 IRS 3A, 3B, and 3C, using IAU
nomenclature.

In addition to the multiplicity of young sources in L1448 IRS 3, 
multiple outflows in the region have been suggested
in previous studies.
\citet{bachiller1990, bachiller1995} suggested an outflow originating from IRS 3B, 
since they detected a redshifted component in the blueshifted lobe of the mm-source outflow
and since IRS 3B is the brightest source at mm wavelengths.
In contrast, \citet{curiel1990} suggested that the outflow was driven by IRS 3A,
based on a spectral index of 0.2, similar to thermal jet models \citep[e.g.][]{reynolds1986}, 
and coincident H$_2$O maser observations.
\citet{davis1995} and \citet{eisloffel2000} presented H$_2$ emission images showing shocked regions.
They suggested up to three outflows in the IRS 3 region from collimated features in
the images.
Several Herbig-Haro objects were also detected and explained
as outflows driven by the three IRS 3 sources \citep{bally1997}.
\citet{wolfchase2000} suggested two outflows from IRS 3A and 3B, based on previous 
studies and their large-scale maps of CO $J=1\rightarrow0$\ emission.
Moreover, \citet{girart2001} reported a well collimated redshifted component detected 
along a line of position angle $110 \arcdeg$ from IRS 3B in SiO $J=2\rightarrow1$ emission.
These previous studies, however, could not clearly show the outflow features 
relative to the sources due to a lack of angular resolution and 
the inherent complexity of the region.

Theoretical studies have suggested that magnetic fields play key roles in the outflows 
of protostars as well as star formation itself.
Observations of magnetic morphology and strength are possible using the Zeeman
effect \citep[e.g.][]{crutcher1999} and linear polarization of dust emission and spectral lines 
\citep[e.g.][]{girart1999}.
\citet{crutcher2004} recently estimated the magnetic strength as well as morphology, from 
linear polarization of dust
emission in three prestellar cores (starless cores) using the Chandrasekhar-Fermi method \citep{chandrasekhar1953}.
In the case of magnetic fields related to protostars (Class 0 sources)
with outflows,
\citet{girart1999} detected linear polarization in the $\lambda=1.3$ mm dust emission and
CO $J=2\rightarrow1$\ spectral line.
Their study was unique for low mass protostars with outflows.

In this paper, we present polarimetric observations in the $\lambda=1.3$ mm continuum and 
CO $J=2\rightarrow1$\ showing two outflows and magnetic fields in the L1448 IRS 3 region, using the BIMA array.
Moreover, \tco~$J=1\rightarrow0$ and \ceo~$J=1\rightarrow0$ observation results are presented.
We argue that the binary system of L1448 IRS 3A and 3B is gravitationally bound,
using these \tco~and \ceo~observations.
 
\section{Observation and data reduction}
\label{sec_obs}

We performed $\lambda=1.3$ mm continuum and CO $J=2\rightarrow1$\ polarimetric observations toward L1448 IRS 3,
using nine of the 10 antennas in
the BIMA array\footnote{The BIMA Array was operated by the Berkeley Illinois
Maryland Association under funding from the National Science Foundation.
BIMA has since combined with the Owens Valley Radio Observatory millimeter
interferometer, moved to a new higher site, and was recommissioned as
the Combined Array for Research in Millimeter-wave Astronomy (CARMA) in 2006.} \citep{welch1996}.
The data were obtained on 2003 October 17, 25, 26, and November 13 in C configuration and
integrated for 9 hours except October 17, which was 3 hours due to weather.
To get $\lambda=1.3$ mm continuum data as well as CO $J=2\rightarrow1$\ spectral line,
a correlator mode with 4 windows of 100 MHz bandwidth each in both sidebands was used.
This mode gives $\sim 4 \textrm{ \kms}$ channel width for the CO $J=2\rightarrow1$\ spectral line.
The first window of the upper side band was centered at the rest frequency 
of CO $J=2\rightarrow1$, 230.538 GHz.
The synthesized beam, obtained through natural weighting, is around $2\farcs5\times4\farcs5$.

0237+288 was used as a secondary flux calibrator as well as a phase calibrator; 
Uranus was the primary flux calibrator.
We set the flux of 0237+288 of all data on the value determined from Uranus 
observed October 25, the best data set.
As the 0237+288 flux is not variable in such a short period (40 days), 
it is arguably the most consistent combination of the data.
The flux of 0237+288 estimated by the primary flux calibrator is 1.33 Jy;
our flux calibration is estimated at a 20\% absolute uncertainty.

Each antenna of the BIMA array has a quarter-wave plate in the front of the linearly 
polarized feed for polarimetric observations.
The quarter-wave plate gives left (L) or right (R) circular polarizations and
the cross correlations (LL, RR, LR, RL) enable the calculation of the Stokes parameters.
To obtain quasi-simultaneous measurements of dual polarizations,
antennas switch to measure L or R circular polarization following a fast Walsh function.
The data are averaged over the Walsh cycle.  
The details on the BIMA polarimetric instrument can be found in \citet{raophd1999}.
In the polarimetry observations, instrumental leakage must be compensated.
The leakage terms are $\sim 5$\% at $\lambda\sim 1$ mm
and constant until quarter-wave plates are reinstalled \citep{raophd1999,rao1998}. 
In addition, they are strongly dependent on frequency.
We used 3C279 data observed on 2003 March 4 at the same frequency (230.538 GHz)
to get the leakage terms.

MIRIAD \citep{sault1995} was used to reduce our data.
First, we applied gains obtained from 
calibrators and constructed models for each observation date.
Data of each observation date were self-calibrated with the model constructed
from their own data.
After combining the individually self-calibrated data, we constructed a
combined model.
Finally, all data were individually self-calibrated again with this 
model and combined to the final result.

\tco~$J=1\rightarrow0$\ ($\nu_{rest}=110.201$ GHz) and \ceo~$J=1\rightarrow0$\ ($\nu_{rest}=109.782$ GHz) 
data were obtained on 2004 April in C configuration of the BIMA array.
These two spectral lines were observed simultaneously with a channel width of
$\sim 1 \textrm{ \kms}$ and synthesized beam size of $\sim 8\arcsec \times 7\arcsec$.
Uranus was used as a primary flux calibrator and 0336+323 as a phase calibrator and
a secondary flux calibrator. The estimated 0336+323 flux was 1.65 Jy.
Again, this flux calibration is estimated at a 20\% absolute uncertainty.

Submillimeter continuum observations of L1448 IRS 3 at 850~$\mu$m were accessed from 
the JCMT data archive. They had originally been observed with SCUBA \citep{holland1999}
on the JCMT on Mauna Kea, during the evenings of 1999 August 28 (6 pointings), 
2000 January 3 (7 pointings), and 2000 February 24 (2 pointings). 
SCUBA was used with the SCUBAPOL \citep{greaves2003} polarimeter, 
which uses a rotating half-wave plate and fixed analyzer. 
The wave plate is stepped through sixteen positions (each offset from the last by $22.5 \arcdeg$) 
and a Nyquist-sampled image (using a 16-point jiggle pattern) is taken at each wave plate 
position \citep{greaves2003}. 
The observations were carried out while chopping the secondary mirror 120 arcsec 
in azimuth at 7 Hz and synchronously detecting the signal, thus rejecting sky emission. 
The integration time per point in the jiggle cycle was 1 second, 
in each of the left and right telescope beams of the dual-beam chop. 
The total on-source integration time per complete cycle was 512 seconds. 
The instrumental polarization (IP) of each bolometer was measured on the planets 
Mars and Uranus. 
This was subtracted from the data before calculating the true source polarization. 
The mean IP was found to be 0.93$\pm$0.27\%. 
The submillimeter zenith opacity for atmospheric extinction removal was determined 
by comparison with the 1.3-mm sky opacity \citep{archibald2002}.

\section{Dust continuum emission}
\label{sec_cont}

\citet{looney2000} revealed three Class 0 sources in this region in the
$\lambda=2.7$ mm continuum with high resolution BIMA observation
and denominated as L1448 IRS 3A, 3B, and 3C (hereafter IRS 3A, 3B, and 3C).
Note that some authors used L1448 N(A), N(B), and NW, respectively 
\citep{terebey1997, barsony1998}.
\citet{ciardi2003} reported a mid-infrared ($10\sim 25~\mu$m) observation of IRS 3A and 3B.
They suggested that IRS 3A and 3B are Class I and Class 0 sources respectively,
based on a comparison of the envelope and central source masses.
On one hand, IRS 3A and 3B could be a ``coeval'' binary system with different central masses and so
be evolving at different rates.
On the other, they may be evolving at a same rate under different environments due to
interaction, a mass flow from one to the other.
Although we do not focus on distinguishing the two cases, we discuss the binarity
(i.e. if the two sources are gravitationally bound) later, which is a basis for the
two ideas.

\begin{figure}
\includegraphics[width=0.8\textwidth, angle=270]{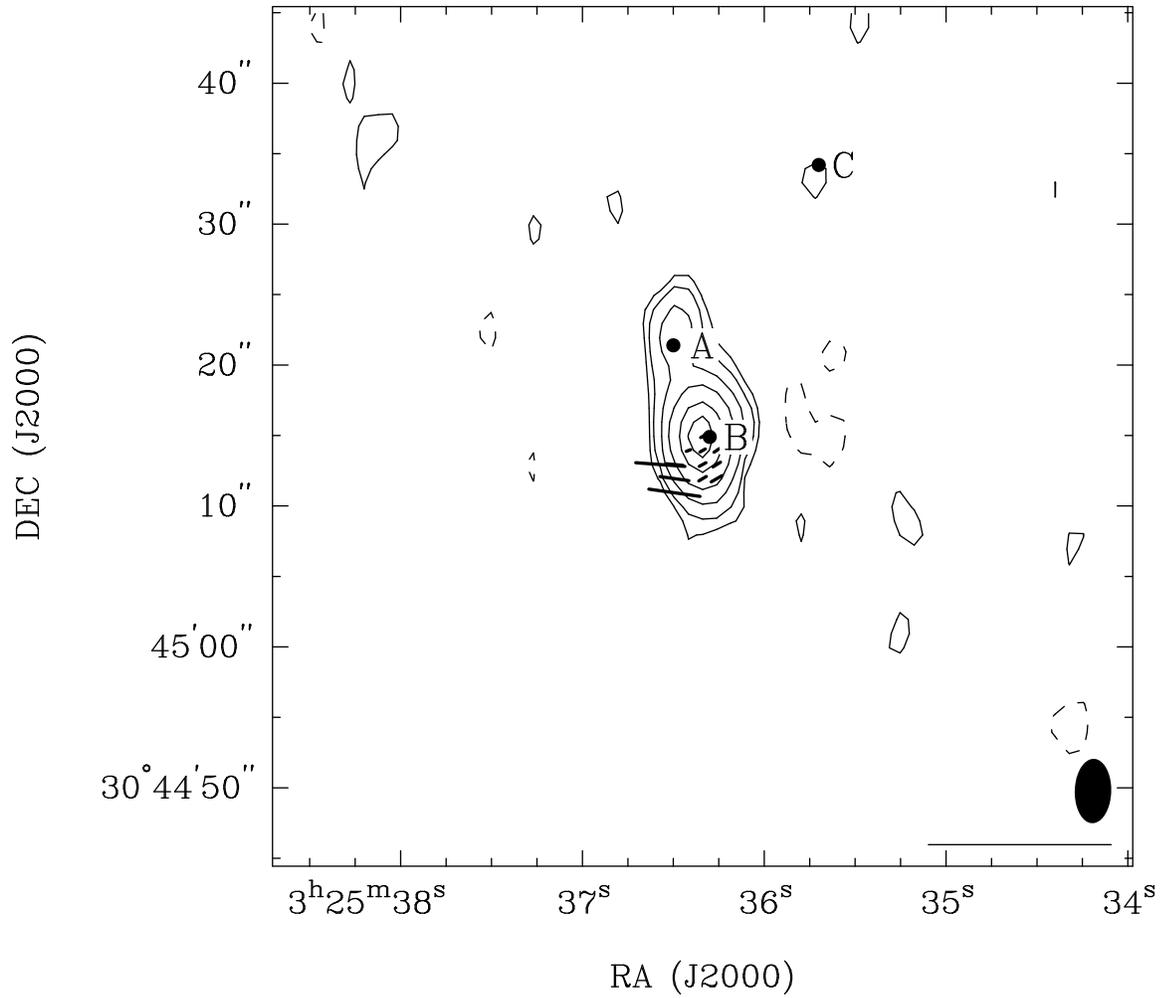}
\caption{$\lambda=1.3 \textrm{ mm}$ continuum map of L1448 IRS 3.
Vectors indicate linear polarization and the symbols at bottom right show
the synthesized beam of $4\farcs6\times2\farcs6$~($\textrm{PA}=-1.6 \arcdeg$) and 100\% polarization scale. 
Contour levels are 3, 5, 10, 20,
40, and 60 times $\sigma=9.4 \textrm{ mJy beam}^{-1}$. \label{contpolmap}}
\end{figure}

Figure \ref{contpolmap} shows the observed $\lambda=1.3$ mm continuum image.
IRS 3A and 3B are distinct, but IRS 3C is marginally detected.
The vectors in Figure \ref{contpolmap} indicate the linear polarization direction,
which will be discussed in \S \ref{sec_magfields}.
The locations of IRS 3A, 3B, and 3C in Figure \ref{contpolmap} are from 
\citet{looney2000}.
Table \ref{sources} summarizes locations, fluxes, and estimated masses of the three 
sources.
To estimate the mass of the circumstellar material (envelopes and disks), 
we assume optically thin dust emission and a uniform envelope 
dust temperature of 35 K,
\begin{equation}
F_{\nu}=B_{\nu}(T_{dust})~\kappa_{\nu}~M_{tot}~D^{-2}
\end{equation}
where $B_{\nu}$ is a black-body intensity of a temperature $T_{dust}$, 
$\kappa_{\nu}$ is a mass absorption coefficient, $M_{tot}$ is
the total mass of gas and dust, and D is the source distance.
We assume a mass absorption coefficient, 
$\kappa_{\nu}=0.005 \textrm{ cm}^2 \textrm{ g}^{-1}$ at $\lambda=1.3$ mm.
The mass absorption coefficient was acquired following a dust emissivity model of
a power law ($\kappa_{\nu}\sim \lambda^{-\beta}$) with $\beta=1$.
Dust emissivity studies of submillimeter wavelengths suggested
$\lambda^{-1}$ dependence in circumstellar disks and dense cores rather than
$\lambda^{-2}$ \citep{weintraub1989,beckwith1990,beckwith1991,looney2003}.

\section{\tco~$J=1\rightarrow0$\ and \ceo~$J=1\rightarrow0$\ observation}
\label{sec_13coc18o}

\begin{figure}
\includegraphics[angle=270, width=0.95\textwidth]{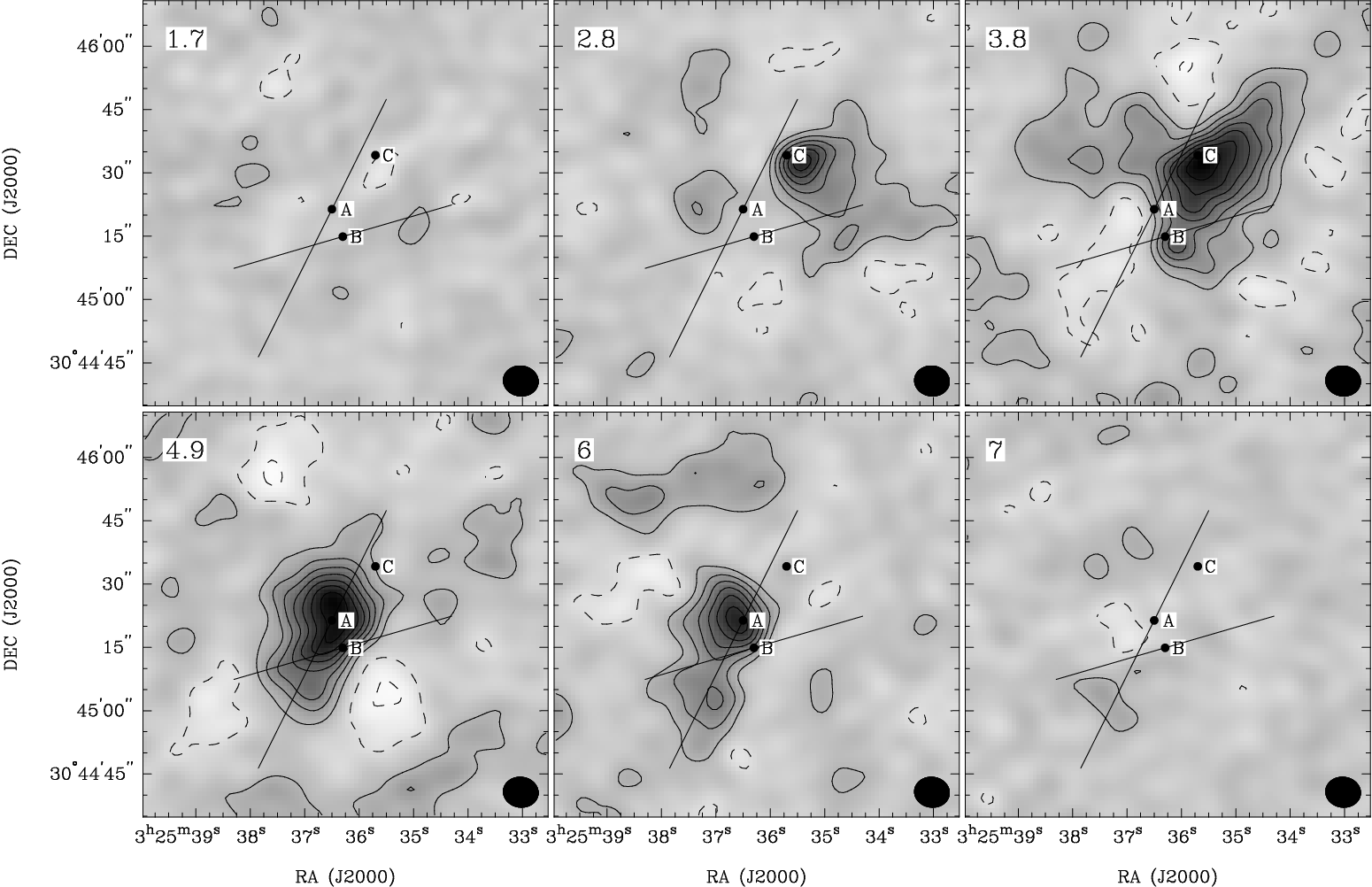}
\caption{$^{13}$CO $J=1\rightarrow0$\ channel maps of L1448 IRS 3. 
Two lines indicate the outflow directions 
that are discussed in \S \ref{sec_outflows}. Source locations and beam size are marked.
The synthesized beam is $8\farcs1\times7\farcs0$ and $\textrm{PA}=82 \arcdeg$.
Contour levels are 3, 7, 11, 15 19, 23,
31, and 35 times $\sigma=76 \textrm{ mJy beam}^{-1}$.   \label{13cochmap}}
\end{figure}

\begin{figure}
\includegraphics[angle=270, width=0.95\textwidth]{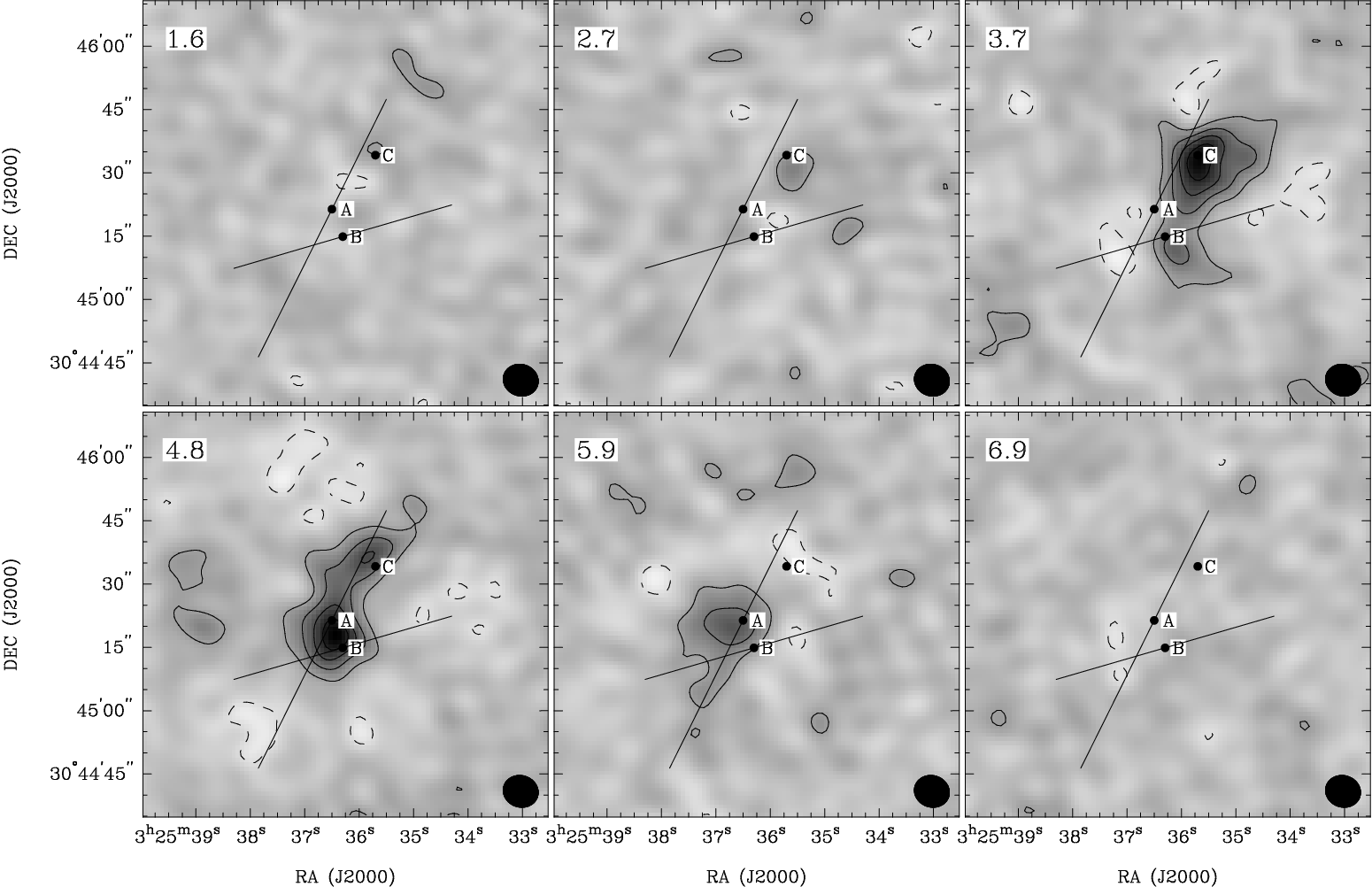}
\caption{C$^{18}$O $J=1\rightarrow0$\ channel maps of L1448 IRS 3. Two lines indicate the outflow directions 
that are discussed in \S \ref{sec_outflows}. Source locations and beam size are marked.
The synthesized beam is $8\farcs2\times7\farcs3$ and $\textrm{PA}=72 \arcdeg$.
Contour levels are 3, 7, 11, and 15 
times $\sigma=76 \textrm{ mJy beam}^{-1}$.\label{c18ochmap}}
\end{figure}

We detect all three sources (IRS 3A, 3B, and 3C) in \tco~and \ceo.
Figures \ref{13cochmap} and \ref{c18ochmap} show the \tco~$J=1\rightarrow0$\ 
and \ceo~$J=1\rightarrow0$ channel maps, respectively.
IRS 3C peaks are in the $3.8 \textrm{ \kms}$ channel of \tco~and $3.7 \textrm{ \kms}$ channel of \ceo.
The emission from IRS 3A and 3B is distributed over different velocity cannels;
IRS 3A is around the $5$ and $6 \textrm{ \kms}$ channels in the two spectral lines, while
IRS 3B is around the $4$ and $5 \textrm{ \kms}$ channels.
This implies that the envelopes of IRS 3A and 3B have a velocity difference 
less than $1 \textrm{ \kms}$.
\citet{terebey1997} have reported a velocity-position diagram of \ceo~$J=1\rightarrow0$
with a comparable angular and spectral resolution to ours, showing these velocity
differences between IRS 3A, 3B, and 3C.
They suggested kinematics of the binary system IRS 3A and 3B, as well as
a rotating system consisting of IRS 3C and the common envelope of IRS 3A and 3B.
However, they did not discuss the physical conditions of the presumed binary system.
We discuss the binary system of IRS 3A and 3B in \S \ref{sec_binary}, 
based on their velocity difference and the two outflows shown in the following section.
Moreover, we estimate the specific angular momentum of the binary system.

Isotopic observations such as \tco~ and \ceo~ are used to trace denser regions and 
to verify their optical depth.
However, we do not follow the standard procedure because the masses of envelopes are 
better estimated using our dust emission data, especially in the case of complicated 
regions like IRS 3.
Also, these isotopes may not trace outflows.
Instead, we use the optical depth results of \citet{bachiller1990}, deduced from 
CO $J=1\rightarrow0$\ and CO $J=2\rightarrow1$, and follow their procedures to
estimate the outflow masses in \S \ref{sec_outflowmass}. 

\section{CO $J=2\rightarrow1$\ observation}
\label{sec_outflows}
\subsection{Bipolar Outflows}
As introduced in \S \ref{sec_intro}, one, two, or up to three outflows have 
been suggested for this region.
\citet{bachiller1990} proposed that an outflow in the east-west direction originates 
from IRS 3, based on  
a redshifted component that was detected in the region of the blueshifted lobe of 
the mm source outflow.
Recently \citet{wolfchase2000} suggested outflows of position angle $150 \arcdeg$ and
$129 \arcdeg$ from IRS 3A and 3B respectively, 
using their large-scale CO $J=1\rightarrow0$\ observation as well as previous studies of
H$_2$ observations and Herbig-Haro objects.
In addition, \citet{girart2001} presented a redshifted SiO component along
a line of position angle of $110 \arcdeg$ from IRS 3B.
However, to date there were no observations with enough angular resolution to clearly
identify outflows with sources.
Here we present high angular resolution BIMA observations to illustrate outflows
in IRS 3.
We reveal two outflows from IRS 3A and 3B but no outflow from IRS 3C, based on 
channel maps and integrated intensity maps.

Figure \ref{cochmap} shows the CO $J=2\rightarrow1$\ channel maps with a velocity range 
from $+29$ to $-36 \textrm{ \kms}$.
The values in the upper left of each panel indicate the channel central velocities 
in units of \kms~and the two lines in each panel show our determined directions of 
the two outflows originating from IRS 3A and 3B.
As introduced in \S \ref{sec_13coc18o}, the $V_{LSR}$ of these sources 
is around $5 \textrm{ \kms}$, which is located at the
boundary between the $+7$ and $+3 \textrm{ \kms}$ channels.

The outflow of IRS 3A is mainly shown in two channels around $V_{LSR}$, 
$+7$ and $+3 \textrm{ \kms}$ channels,
with symmetric features cross the central source.
Some redshifted components of the IRS 3A outflow also appear 
in the $+15$ and $+11 \textrm{ \kms}$ channels. 
The feature might be just an elongated cloud. However, the \tco~and \ceo~maps
indicating ambient clouds do not show such features. 
In addition, the facts that two channels of $+7$ and $+3 \textrm{ \kms}$ show a very similar
shape to each other, that two redshifted channels of $+15$ and $+11 \textrm{ \kms}$ have 
blobs on both sides of IRS 3A, and that H$_2$ emission observations
reveal a string of H$_2$ knots in a consistent direction \citep{davis1995, eisloffel2000},
strongly suggest that it is an
outflow nearly perpendicular to the line of sight, originating from IRS 3A.
The position angle of the IRS 3A outflow is $155 \arcdeg$.

As shown in Figure \ref{cochmap}, the outflow from IRS 3B appears 
from the $+23 \textrm{ \kms}$ channel, an end channel of a redshifted lobe, 
along the position angle of $105 \arcdeg$.
This position angle is consistent with the \citet{girart2001} estimate of $110 \arcdeg$.
The redshifted lobe is clearly seen from the $+23$ to $+11 \textrm{ \kms}$ channels and overlaps
with the southern lobe originating from IRS 3A in the $+7$ and $+3 \textrm{ \kms}$ channels.
Note that the blueshifted channels after $+3 \textrm{ \kms}$ look complicated with many blobs.
This can be explained by the overlap with the blueshifted 
lobe of the mm source outflow, located $70''$ southeast.
Indeed, the velocity range of the mm source's blueshifted outflow lobe is 
consistent with these channels \citep[see Fig.10 in][]{bachiller1990}. 
Due to the complexity in blueshifted channels, the outflow direction is deduced
from the redshifted channels first.
However, we can still see the blueshifted components of the IRS 3B outflow 
up to $-30 \textrm{ \kms}$ along the $105 \arcdeg$ position angle.
In addition, a string of three blobs along the outflow direction is shown in the
$-5.2 \textrm{ \kms}$ channel.
The blobs in the $+15$ and $+11 \textrm{ \kms}$ channels can be the opposite components 
of these three blobs.

Figure \ref{com0map} is an integrated intensity map of CO $J=2\rightarrow1$\ that more clearly shows the
outflows.
The red contours present a velocity range from $+25$ to $+9 \textrm{ \kms}$, black contours
from $+9$ to $+1 \textrm{ \kms}$, and blue contours from $+1$ to $-32 \textrm{ \kms}$.
Again, the blue contours appear with several blobs due to
the effect of the overlapped blueshifted lobe of the mm source outflow.
This integrated intensity map confirms 
two outflows originating from IRS 3A and 3B as suggested from channel maps.
Black contours in Figure \ref{com0map} show an outflow from IRS 3A and
especially the redshifted and blueshifted lobes around IRS 3B are clearly seen.

\subsection{Mass, Momentum, and Energy}
\label{sec_outflowmass}

\citet{bachiller1990} showed that the outflow regions in IRS 3 are optically thin in CO,
using CO $J=2\rightarrow1$\ and $J=1\rightarrow0$\ lines instead of the CO isotopes 
\tco~or \ceo, since the
outflow regions are not dense enough to be traced by these isotopic observations.
In addition, they estimated the excitation temperature of CO. 
As described in \citet{bachiller1990},
the CO column density can be estimated from the integrated intensity,
assuming optically thin emission and level populations
in local thermal equilibrium at the excitation temperature $T_{21}$,
\begin{equation}
\frac{N(CO)}{cm^{-2}} = 1.06 \times 10^{13}~T_{21}~exp(16.5/T_{21})
\int T_R(2-1)~d\Big(\frac{v}{km~s^{-1}}\Big).
\end{equation}
To estimate the masses of the lobes, we assume a typical abundance of CO, 
$\textrm{CO/H$_2$}=\textrm{x(CO)} \sim 10^{-4}$,
optically thin CO emission in outflow regions,
and a CO excitation temperature of 11 K, as reported by \citet{bachiller1990}.
These assumptions give the mass estimate equation,
\begin{eqnarray}
M_{lobe} &=& 2m_H~A_{lobe}~10^4~N(CO) \nonumber \\
&=& 2m_H~5.23\times10^{-5}~\frac{\lambda^2 D^2}{2k}~\int \frac{F_{lobe}(CO~2-1)}{Jy}~
d\Big(\frac{v}{km~s^{-1}}\Big) \nonumber \\
&=& 3.22\times10^{-6}~\textrm{\msun}~ \Big(\frac{D}{250~pc}\Big)^2~ 
\int \frac{F_{lobe}(CO~2-1)}{Jy}~d\Big(\frac{v}{km~s^{-1}}\Big)~
\end{eqnarray} 
where $A_{lobe}=\Omega D^2$ is the lobe area and $F_{lobe}$ is the total lobe flux density.
The mass estimates are summarized in Table \ref{estm_outflow} with momentum and kinetic
energy.
Note that the errors tabulated are only statistical errors.
Moreover, important uncertainties like CO opacities and
unknown 3-D geometry of outflows affect these
outflow mass estimates \citep{lada1985, bachiller1990}.
In the case of the blueshifted lobes, there is also the uncertainty from the 
contribution of the mm source outflow. 

We also need to consider that interferometric observations resolve out flux from extended 
structures.
The missing flux makes mass estimates of extended structures difficult and underestimated.
On the other hand, interferometry is a powerful technique that can reveal small structures
overlapped with large-scale emission.
In this case, L1448 IRS 3 is overlapped with the large blueshifted lobe of the L1448-mm 
outflow.
Therefore, we have the advantage of minimizing the L1448-mm outflow contamination as well as
the disadvantage of losing the flux of extended features.
The {\it uv} coverage of our observations allows us to recover flux up to $15\arcsec$ structures.
The missing flux is less significant in elongated structures because it also
depends on the size-scales of the minor axis.

Compared to the single dish observations (beam size $\sim 12\arcsec$) 
of \citet{bachiller1990}, there is no significant missing flux in the
redshifted wing ($\gtrsim 10 \textrm{ \kms}$); the flux is consistent within the uncertainties.
The low velocity components ($0$ to $10 \textrm{ \kms}$) are dominated
by the ambient cloud \citep{bachiller1990}.
Since our two channels in that velocity range still show outflow features consistent with 
the redshifted channels (see Fig. \ref{cochmap}), we argue that the majority of missing flux comes
from the ambient cloud emission.
From this point of view, the missing flux is a large advantage as it avoids contamination
with the ambient cloud emission rather than the disadvantage of losing flux.
On the other hand, the channel with $-1.1 \textrm{ \kms}$ central velocity may experience
relatively large flux loss, as the IRS 3B outflow feature disappears or is indistinct
from the extended, and mostly resolved-out, blueshifted lobe of the L1448-mm source.
The missing flux in this channel would cause an underestimation of the IRS 3B outflow mass 
but does not significantly affect our interpretations.
We discuss the effects caused by the missing flux in this channel later in related sections.
Overall, although the true size of the emission is unknown, we conclude
that the missing flux probably does not significantly affect our results.

We estimate the masses of the northern and the southern lobes of the IRS 3A outflow
as $0.70\times10^{-3}$ and $1.12\times10^{-3} \textrm{ \msun}$, respectively.
Considering that the southern lobe is overlapped with a portion of a redshifted
lobe of an outflow from IRS 3B, we can regard $0.42\times10^{-3} \textrm{ \msun}$
(the mass difference of the two lobes) as coming from IRS 3B.
This assumes that the two lobes from an outflow have similar masses.
As a result, the northern and the southern lobes of IRS 3A outflow 
have $\sim 0.70\times10^{-3} \textrm{ \msun}$ each.

The redshifted and blueshifted lobes of the outflow originating
from IRS 3B are well distinguished in the velocity range $+25$ to $-32 \textrm{ \kms}$.
The outflow spans the blue, black, and red contours in Figure \ref{com0map}.
The blueshifted lobe of this outflow has $0.98\times10^{-3} \textrm{ \msun}$, which is comparable to
the $0.75\times10^{-3} \textrm{ \msun}$ of the redshifted lobe as a combination of the 
red contours ($0.33\times10^{-3} \textrm{ \msun}$) 
and a portion of the black contours ($0.42\times10^{-3} \textrm{ \msun}$).
The difference is reasonable, 
considering the blueshifted regions are contaminated by components of the mm source outflow,
probable flux loss in the $-1.1 \textrm{ \kms}$ channel, and flux gain in other blueshifted channels.
In summary, the redshifted lobe of the IRS 3B outflow has $0.75\times10^{-3} \textrm{ \msun}$ and the 
blueshifted lobe has $0.98\times10^{-3} \textrm{ \msun}$, comparable within the uncertainties.

Estimates of momentum and kinetic energy of each component are also shown in 
Table \ref{estm_outflow}.
We use $V_{LSR} = 5 \textrm{ \kms}$ and do not apply the inclination factors, 
which are $1/cos\theta$ for momentum and $1/cos^2\theta$ for kinetic energy. 
Here $\theta$ is the inclination angle from the line of sight. 
When calculating the momentum and kinetic energy, we assume that components in each channel 
have the central channel velocity.
Comparing momentum and energy of each lobe of the two outflows,
the southern lobe of the IRS 3A outflow and the eastern lobe of the IRS 3B
outflow have lower momentum and kinetic energy than their opposite lobes.
Although the contamination of the L1448-mm outflow 
in the blueshifted lobe of the IRS 3B outflow
and the nearly perpendicular aspect of the IRS 3A outflow to the line of sight make the
interpretation difficult,
it is probable that the outflows from IRS 3A and 3B interact
in the overlapped region because the kinetic energy difference is distinct even when considering
a portion of the blueshifted lobe of IRS 3B to be the same mass as the redshifted lobe.
Due to the collision of two outflows, the kinetic energy
would be reduced. The fact that there is no blueshifted opponent of the small 
redshifted blobs of the
IRS 3A outflow in the $+15 \textrm{ \kms}$ channel supports this idea as well. 
Besides, the heated clump, which \citet{curiel1999} presented near IRS 3 in
observations of the NH$_3~(J, K)=(1, 1)$ and $(2, 2)$ inversion transitions,
is located in the overlapped region of the two outflows from IRS 3A and 3B.
Although they argued that the heated clump would be a part of a larger heated region
because IRS 3 is at the edge of their field,
it may present the interaction of the two outflows. 
In addition, the fact that the redshifted component of the IRS 3B outflow detected in 
SiO $J=2\rightarrow1$ over the interaction region has relatively low velocity compared to
the blueshifted component \citep{girart2001}, also supports interaction. 
Based on these considerations, we also suggest that IRS 3B is closer than IRS 3A
because this deployment can reproduce the interaction. 
This is pointed out again later, in the CO $J=2\rightarrow1$\ linear polarization of \S \ref{sec_magfields}.

\subsection{Velocity, Inclination, and Opening}
Based on the mass of the outflow lobes and an assumed mass loss rate,
we can check whether or not the IRS 3A outflow is nearly perpendicular.
If we assume a mass loss rate of  $6\times10^{-7}\textrm{ \msun~yr}^{-1}$ from the 
two outflow lobes\footnote{
This is consistent with previous studies suggesting  
that massive young stars lose mass up to $10^{-3}\textrm{ \msun~yr}^{-1}$
and low mass stars down to $10^{-9}\textrm{ \msun~yr}^{-1}$
\citep[e.g.][]{kim2006,wu2004,bontemps1996}.
In order to obtain a reasonable outflow velocity ($\sim 10 \textrm{ \kms}$), 
we assume a mass loss rate of $6\times10^{-7}\textrm{ \msun~yr}^{-1}$.},
the age of the outflow would be $\sim 2300$ years and the proper
velocity of the outflow would be $\sim 10 \textrm{ \kms}$ since the outflows extend to $20''$ 
($5000$ AU at $250$ pc).
As the channel width is $4 \textrm{ \kms}$ and the $V_{LSR}$ is on the boundary of
the two channels showing the outflow feature,
the inclination including opening angle must be less than $22 \arcdeg$ from the plane of
the sky; the IRS 3A outflow is nearly perpendicular to the line of sight.

The inclination angle of IRS 3B can be estimated from the velocity features detected.
Figure \ref{vpdiag} presents a velocity-position diagram cut along the outflow direction
of position angle $105 \arcdeg$ from IRS 3B.
Both redshifted (east) and blueshifted (west) lobes are divided into two components: one 
accelerating from the source and the other with a constant velocity.
Using Figures \ref{vpdiag} and \ref{cochmap}, the 
accelerating portion up to $+23 \textrm{ \kms}$ (channel central velocity) and the constant
part at $+3 \textrm{ \kms}$ are in the eastern (redshifted) region.
Similarly, the accelerating portion up to $-30 \textrm{ \kms}$ and the constant part at
$-5 \textrm{ \kms}$ are in the western (blueshifted) region.
These velocities are $+18$, $-2$, $-10$, and $-35 \textrm{ \kms}$ in the IRS 3B rest frame 
($V_{LSR}=+5 \textrm{ \kms}$).
If the missing flux in the $-1.1 \textrm{ \kms}$ channel is part of the constant component, then
arguable $-6 \textrm{ \kms}$ in the IRS 3B rest frame is a better extreme constant velocity.
However, since the small difference in velocity will not significantly change the results derived below
and since it is strongly dependent on the assumed missing flux, 
we use $-10 \textrm{ \kms}$ in the IRS 3B rest frame  as the velocity of the constant 
velocity component in blueshifted region.

The constant and accelerating features are best explained 
by two possible geometric outflow effects,
although an outflow with various velocity components is also a possible explanation.
One is the geometric effect caused by precession and the other by a trumpet-shaped outflow.
The precession of the IRS 3B outflow-- the side of the redshifted lobe is precessing
toward the observer and the side of the blueshifted lobe is away from the observer, 
would give the detected map features.
In other words, the redshifted or blueshifted components further from the central source 
are older components emitted when the inclination was smaller than now.
Therefore, the outflow is observed as accelerating away from the source, 
even assuming a constant outflow velocity.
A trumpet-shaped outflow can also give the detected features.
The ``trumpet'' outflow has different angles with respect to the line
of sight along the redshifted or blueshifted lobes.
These different angles can give the accelerating feature depending on the outflow inclination.

The nice aspect of the ``trumpet'' outflow is that we can estimate outflow parameters
such as velocity, inclination angle, and opening angle, based on the observed data.
The opening angle is assumed as the angle on the end of the outflow ``trumpet''.
Therefore, the velocity difference between accelerating and non-accelerating features 
of the redshifted or blueshifted lobe is coming from the opening angle.
In addition to the velocity ($v$), the inclination angle ($\theta_i$), 
and the opening angle ($\theta_o$),
we adopt the velocity difference ($\Delta v$) and the opening angle difference ($\Delta\theta_o$) 
between the redshifted and blueshifted lobes.
Note that the inclination angle is measured from the line of sight and that 
the opening angle is half of the outflow opening.
As mentioned in the previous section, since the redshifted lobe of the IRS 3B outflow is
likely to be interacting with the southern lobe of the IRS 3A outflow, 
a velocity difference needs to be included.
The reason for applying the opening angle difference is that the side of the redshifted lobe has
components in a blueshifted channel with a central velocity of $+3 \textrm{ \kms}$ (Fig. \ref{cochmap}),
while the side of the blueshifted lobe does not.
We define the velocity difference as $\Delta v=v_{blue}-v_{red}>0$ ($v=v_{blue}$) and
the opening angle difference as $\Delta\theta_o=\theta_{o,red}-\theta_{o,blue}>0$ 
($\theta_o=\theta_{o,blue}$).
The blueshifted lobe is expected to have a higher velocity and a narrower opening angle
than the redshifted lobe.
Using Bayesian statistics, we determine the most likely parameter combinations
to explain the observed velocity features,
\begin{eqnarray}
P(\{parameters\}|\{v'_i\}) &=& \frac{P(\{v'_i\}|\{parameters\})}{P(\{v'_i\})}~
P(\{parameters\}), \nonumber \\
\textrm{where}~~~ \{parameters\} &=& \{v, ~\theta_i, ~\theta_o, ~\Delta v, ~\Delta\theta_o \}.
\end{eqnarray}
The $\{v'_i\}$ is a set of four extreme values of the observed line-of-sight velocities 
in the outflow lobes with respect to the IRS 3B rest frame ($v_1$, $v_2$, $v_3$, and $v_4$),
so the evidence term, $P(\{v'_i\})=1$.
In addition, since we do not have any preference to choose the five parameters, 
we assume that the prior probability densities, $P(\{parameters\})$, are uniform. 
However, note that the opening angle should be less than $45 \arcdeg$; otherwise,
we would observe redshifted components on the side of the blueshifted lobe as well as
the blueshifted components on the side of the redshifted lobe.
For the likelihood, $P(\{v'_i\}|\{parameters\})$, we choose a probability density 
having a constant value in the channel width ($4 \textrm{ \kms}$) and 
exponentially decreasing outside of the channel width.

We found the parameters giving the maximum posterior probability,
after taking into account central velocities of four-end channels showing outflow features
in the IRS 3B rest frame, $\{v'_i\}=\{+18,-2,-10,-35\} \textrm{ \kms}$.
The maximum posterior is obtained when the four velocities estimated from
the five parameters are in the flat-top regions of each channel.
Note that we adopted the likelihood (indicating channels) of functions 
having a constant value (flat-top) in the channel width ($4 \textrm{ \kms}$) and 
exponentially decreasing outside.
To explore the parameter space, the Metropolis-Hastings method \citep{mackay2003}
was used; we obtained a few 
hundred thousand samples with the maximum posterior probability through one million trials.
Since we use flat-top shaped channel functions for the likelihood, 
the parameters are largely distributed.
For example, the velocity distribution peaks at around $40 \textrm{ \kms}$ and quickly drops, 
but some cases have even a few hundred \kms.
The inclination angle and the opening angle are distributed in $30\sim 80 \arcdeg$ and 
$5\sim 45 \arcdeg$, respectively.

However, as these parameters are not independent of each other,
we can narrow the acceptable range of parameters.
Figure \ref{fig_para} shows the velocity (solid circles), the inclination angle (solid squares), 
and the velocity and opening angle differences (open circles and open triangles, respectively)
versus the opening angle.
The opening angle is also plotted (solid triangles) to compare with the other parameters.
For the plot, samples are divided by $5 \arcdeg$ bins of the opening angle and each bin has 
a few tens of thousand samples.
The data points in Figure \ref{fig_para} are average values of the samples in each bin,
and the error bars present their standard deviation.
The dotted line without data points indicates the opening angle plus the opening angle difference,
in other words, the redshifted opening angle.
Note that small opening angles with relatively large velocities and inclination angles
or large opening angles with small velocities and inclination angles 
give the four observed line-of-sight velocities.
However, opening angles larger than $26 \arcdeg$ are rejected because they have
the $18 \textrm{ \kms}$ component with the $-10 \textrm{ \kms}$ component on the same side, which is not consistent
with our observation (Fig. \ref{vpdiag}).
Too small opening angles ($< 8 \arcdeg$) are also not likely due to the too large velocities required.
As a result, we can constrain the opening angle to 
$8\arcdeg<\theta_o<26\arcdeg$ and the other parameters to the values following the opening angle;
for examples, $100 \gtrsim v \gtrsim 40 \textrm{ \kms}$ and $75\arcdeg \gtrsim \theta_i \gtrsim 50\arcdeg$.
Recent Spitzer Space Telescope (SST) observations have suggested that an opening angle
of $\sim 20\arcdeg$ is preferred \citep{tobin2007}.
In that case, we constrain the inclination angle to $\sim 57 \arcdeg$, 
which is consistent with the SST observational results within uncertainties,
and the velocity to $\sim 45 \textrm{ \kms}$.
These parameters give the age of the IRS 3B outflow detected in our field-of-view 
as $\sim 600 \textrm{ years}$.

\section{Binary system of IRS 3A and 3B}
\label{sec_binary}
The velocity difference between IRS 3A and 3B detected in the \tco~$J=1\rightarrow0$\ 
and \ceo~$J=1\rightarrow0$\ observations can be understood as an orbiting binary system.
This also supports the interaction of the two outflows from IRS 3A and 3B.
Here we introduce a kinematical constraint for a Class 0 binary system.

When denominating velocities of the two clouds with respect to the center of mass of
the binary system as $v_A$ ($>0$) and $v_B$ ($<0$) and
the components of the velocities in the line-of-sight plane of the IRS 3A and 3B
as $v_{A\parallel}$ and $v_{B\parallel}$,
the projected velocities on the line of sight are $v_A'=v_{A\parallel} sin\theta$ and
$v_B'=v_{B\parallel} sin\theta$ (Fig. \ref{fig_sbinary}).
Note that 
the vertical velocity components to the plane
are $v_{A\perp}$ and $v_{B\perp}$, and so
the proper motion velocities are indicated as $(v_{A\parallel}^2 cos^2\theta + v_{A\perp}^2)^{1/2}$ and
$(v_{B\parallel}^2 cos^2\theta + v_{B\perp}^2)^{1/2}$.
Similarly, the projected semimajor axis is $a'= asin\theta\approx7''\times250~\textrm{pc}
=1750~\textrm{AU}$.
These projected velocity difference (here assuming $0.8 \textrm{ \kms}$) and the projected
semimajor axis ($\sim 1750$ AU) allow the estimation of the orbiting period ($P$) of 
the binary system,
\begin{eqnarray}
|v_A| = \sqrt{v_{A\parallel}^2+v_{A\perp}^2} = \frac{2\pi a_A}{P} &>& |v_{A\parallel}| \nonumber \\ 
|v_B| = \sqrt{v_{B\parallel}^2+v_{B\perp}^2} = \frac{2\pi a_B}{P} &>& |v_{B\parallel}|. \nonumber 
\end{eqnarray}
Multiplying by $sin\theta$, they become
\begin{eqnarray}
|v_{A\parallel}|sin\theta &<& \frac{2\pi a_Asin\theta}{P} \nonumber \\
|v_{B\parallel}|sin\theta &<& \frac{2\pi a_Bsin\theta}{P} \nonumber
\end{eqnarray}
and adding each side gives
\begin{equation}
|v_A'| + |v_B'| < \frac{2\pi a'}{P} \label{eq_p}.
\end{equation}
In equation (\ref{eq_p}), since $v_B'$ is negative,
the left hand side is the projected velocity difference, $0.8 \textrm{ \kms}$.
Using the projected semimajor axis ($1750 \textrm{ AU}$), we obtain
an upper limit of the orbiting period, $\sim 6.54\times10^4$ years.
Note that it is much longer than the age ($\sim 2300 \textrm{ years}$) of the IRS 3A outflow,  
which is obtained assuming the mass loss rate,
and the estimated age ($\sim 600 \textrm{ years}$) of the IRS 3B outflow.
Furthermore, since the masses of the two clouds were estimated as $0.21$ and $1.15
\textrm{ \msun}$ in \S \ref{sec_cont} and Table \ref{sources},
we can also estimate an upper limit of the semimajor axis from the Kepler's third law,
\begin{equation}
\frac{a}{AU} = \bigg( \Big(\frac{m_A + m_B}{\textrm{\msun}}\Big) \Big(\frac{P}{yr}\Big)^2 
\bigg)^{1/3}.
\end{equation}
The estimated masses and the orbiting period give a semimajor axis of $\sim 1800$ AU,
a slightly larger value than the projected semimajor axis.
In this case, the $\theta = sin^{-1}(a'/a)$ is about $76 \arcdeg$ (refer to Fig. \ref{fig_sbinary}), which
means that the considered velocity difference ($0.8 \textrm{ \kms}$) is acceptable 
for a gravitationally bound binary system.
Note that the unconsidered central luminous sources can increase the semimajor axis
by 26\% ($=2^{1/3}$) when they are assumed to have identical masses to the
circumstellar material.
This increases the probability of a gravitationally bound binary system.
If the velocity difference were estimated as $0.5 \textrm{ \kms}$ from higher spectral
resolution observations, the upper limit of the orbiting period would be 
$1.05 \times 10^5$ years, the semimajor axis would be $2500$ AU, and 
the $\theta = sin^{-1}(a'/a)$ would be $44 \arcdeg$.
As described above, the total mass and the projected velocity difference
and semimajor axis give a kinematical, gravitationally bound constraint 
on apparent binary systems.
Since the projected velocity difference is within values for a binary system,
we conclude that
the IRS 3A and 3B sources are likely to be gravitationally bound.
Observations with higher spectral resolution will give a better constraint. 

The angular momentum of the binary system is also noteworthy.
We estimate the specific angular momentum (angular momentum per unit mass)
of this binary system using
the projected velocity components, the projected distance, and the mass ratio from 
the mass estimates of the continuum emission at $\lambda=1.3$ mm.
This estimate has uncertainties caused by the ambiguities of velocities of 
the line of sight as well as proper motions.
The ambiguity of the line of sight comes from the broad channel width of our observation.
If we can remove the ambiguity using higher spectral resolution observation, the
estimate would be a lower limit of the specific angular momentum.
The value is $\sim 3 \times 10^{20}$ cm$^2$ s$^{-1}$, similar to the upper limit 
of the specific angular
momentum of binaries and to the lower limit of molecular clouds in Taurus 
\citep{simon1995,goodman1993}.

\section{Magnetic Fields}
\label{sec_magfields}

Linear polarization is marginally detected in both the $\lambda=1.3$ mm
continuum and CO $J=2\rightarrow1$\ spectral line.
Vectors around IRS 3B in Figure \ref{contpolmap} present polarization detected
in the $\lambda=1.3$ mm continuum.
Since polarization of dust emission is perpendicular to the magnetic field,
the magnetic field is expected in the north-south direction around IRS 3B.
This is consistent with the large scale magnetic field observed by SCUBA at
$\lambda=850~\mu$m shown in Figure \ref{scubamap}.
Note that the vectors in Figure \ref{scubamap} also indicate linear polarization
and the direction around IRS 3B is east-west like the $\lambda=1.3$ mm continuum data in 
Figure \ref{contpolmap}.
Toward the center of IRS 3B, weaker linear polarization is detected at both wavelengths.
The polarization fractions are around 5\% in our BIMA 1.3 mm continuum and
around 2\% in the SCUBA data.
This smaller polarization fraction of the SCUBA data is from
the larger beam size of SCUBA smearing out the linear polarization.

Linear polarization of the CO $J=2\rightarrow1$\ emission, tracing the outflows, 
was detected in patches of the overlapped region of
the southern lobe of the IRS 3A outflow and the redshifted lobe of
the IRS 3B outflow.
Figure \ref{copolmap} is an intensity map of two channels combined in velocity 
from $+1$ to $+9 \textrm{ \kms}$.
Vectors present linear polarization directions and two lines on IRS 3A and
3B indicate outflow directions.
According to the Goldreich-Kylafis effect \citep[e.g.][]{kylafis1983dec},
linear polarization
of spectral lines can be either parallel or perpendicular to magnetic
field, depending on the relation between line of sight, magnetic field
direction, and velocity gradient.
Since the polarization was detected in only a few small patches
located in the overlapped region of the IRS 3A and 3B outflows,
it is hard to define the morphology of the magnetic fields.
However,
as the vectors are likely parallel or perpendicular to the IRS 3B outflow,
we suggest that the polarization comes from the IRS 3B outflow.
At the same time, this suggestion implies that the magnetic field
may be perpendicular or parallel to the outflow\footnote{
\citet{girart1999} detected CO $J=2\rightarrow1$\ polarization perpendicular to dust polarization
in NGC 1333 IRAS 4A and interpreted the spectral line polarization as parallel to the magnetic field.
However, polarization in dust continuum and CO $J=2\rightarrow1$\ may not indicate the same 
magnetic field because dust continuum and CO $J=2\rightarrow1$\ trace different density regions;
the magnetic field direction at the central core may change radially
from an hour glass morphology \citep[e.g.][]{fiedler1993}. In addition, detected polarization directions in our
dust emission and CO $J=2\rightarrow1$\ data are likely to be parallel around the IRS 3B center.
Therefore, we cannot define the magnetic field direction as either perpendicular or parallel to
the CO $J=2\rightarrow1$ polarization here.}.
Although we cannot distinguish between the two,
the parallel magnetic field can be from a large scale magnetic field
morphology (hour glass morphology) widely accepted in forming protostars, and
the perpendicular magnetic field can be from a helical structure extended from a toroidal 
magnetic field suggested by some theories \citep[e.g.][]{ostriker1997}.
The fraction of linear polarization is around 6 to 15\%.

Based on both dust polarization and CO $J=2\rightarrow1$\ polarization,
we suggest a unified magnetic field morphology related to the disk and outflow
structures in forming protostars.
The magnetic field inferred from the dust emission, perpendicular to outflow, may
show a toroidal magnetic field around a circumstellar disk 
and the magnetic field inferred from CO $J=2\rightarrow1$\ may
present a large scale morphology parallel to the outflow or helical structure perpendicular
to the outflow.
As discussed in \S \ref{sec_outflows}, we can also suggest that the IRS 3B source is closer, 
because the polarization appears to be from IRS 3B.
Polarization of the farther source (IRS 3A) is harder to detect due to the
foreground cloud (IRS 3B).

\section{Summary and Discussion}
We present CO $J=2\rightarrow1$, $^{13}$CO $J=1\rightarrow0$\ 
and C$^{18}$O $J=1\rightarrow0$\ observations, 
and $\lambda=1.3$ mm continuum and CO $J=2\rightarrow1$\
polarimetric observations.
IRS 3A and 3B are distinctly detected 
with mass estimates of $0.21$ and $1.15 \textrm{ \msun}$ respectively at $\lambda=1.3$ mm, but
IRS 3C is marginally detected with upper mass limit of $0.03 \textrm{ \msun}$ (\S \ref{sec_cont}).
The ambient velocities of IRS 3A, 3B, and 3C are estimated as $5.5$, $4.5$, and 
$3.5 \textrm{ \kms}$, respectively, from \tco~$J=1\rightarrow0$\ and \ceo~$J=1\rightarrow0$\ channel maps
(\S \ref{sec_13coc18o}). 
The two close sources, IRS 3A and 3B, have a velocity difference less than
$1 \textrm{ \kms}$;
the difference is a kinematical constraint on a gravitationally bound binary system.
Moreover, we estimated the specific angular momentum of the binary system as
$\sim 3\times10^{20}~\textrm{cm}^2~\textrm{s}^{-1}$, similar to the upper limit of binaries
and to the lower limit of molecular clouds in Taurus (\S \ref{sec_binary}).

We present CO $J=2\rightarrow1$\ observations showing two outflows, one each from IRS 3A and 3B (\S \ref{sec_outflows}).
The outflow driven by IRS 3A has $\textrm{PA} = 155 \arcdeg$ and is nearly perpendicular to the line
of sight, while the outflow by IRS 3B has $\textrm{PA}=105 \arcdeg$.
In addition, we posit that the two outflows are interacting in the southern lobe of
the IRS 3A outflow and the redshifted lobe of the IRS 3B outflow, based on a
comparison of the kinetic energies of lobes.
Coupled with the fact that
the linear polarization detected in CO $J=2\rightarrow1$\ is likely to come from the IRS 3B outflow,
IRS 3B is closer than IRS 3A.
We also detected that the IRS 3B outflow has accelerating and non-accelerating features in the 
velocity-position diagram, which is interpreted as either a precessing outflow or a ``trumpet'' outflow.
Assuming a ``trumpet'' outflow of IRS 3B rather than its precession, 
we estimated the velocity, inclination angle, and the opening angle, using Bayesian statistics.
The velocity and the inclination angle are constrained between $100$ and $40 \textrm{ \kms}$ 
and between $75\arcdeg$ and $50\arcdeg$, respectively, as the opening angle between $8\arcdeg$ and $26 \arcdeg$.
Furthermore, using an opening angle of $\sim 20\arcdeg$ from Spitzer Space Telescope observations, 
the velocity and the inclination angle of the IRS 3B outflow
are $\sim 45 \textrm{ \kms}$ and $\sim 57\arcdeg$.

Linear polarization in both the $\lambda=1.3 \textrm{ mm}$ continuum and CO $J=2\rightarrow1$\ spectral line is 
marginally detected around the center and outflow of IRS 3B, respectively (\S \ref{sec_magfields}).
The dust emission polarization gives a magnetic field perpendicular to the
outflow, which may be a toroidal magnetic field parallel to the circumstellar disk. 
In contrast, the spectral line polarization suggests either a perpendicular or 
a parallel magnetic field to the IRS 3B outflow.
To determine the relation between the magnetic field and the outflow direction,
more sensitive polarimetric observations are required.

The L1448 IRS 3 is an excellent region to study star formation.
The IRS 3A outflow, nearly perpendicular to the line of sight, enables the
study of the disk and outflow structures in protostar systems more easily, since
it shows the profile projected in the plane of the sky. 
In addition, the binary system of IRS 3A and 3B having
two outflows in quite different directions gives an opportunity to study
the interaction between two sources as well as constrain binary system formation
itself.
Finally, more sensitive polarimetric observations will provide clues on the
connection between outflows and magnetic fields.

\acknowledgments
The authors acknowledge support from the Laboratory for
Astronomical Imaging at the University of Illinois and NSF AST 0228953.
W. K. thanks Ben Wandelt for discussions on Bayesian statistics and
an anonymous referee for helpful comments.
The JCMT Archive project is a collaboration between the Canadian Astronomy
Data Centre (CADC), Victoria and the James Clerk Maxwell Telescope
(JCMT), Hilo. Funding for the CADC's JCMT Archive effort is provided by
the National Research Council of Canada's (NRC) to the Herzberg Institute
of Astrophysics. 

Facilities: \facility{BIMA}, \facility{JCMT (SCUBA)}

\bibliographystyle{apj}                       %
\bibliography{kwon}

\clearpage
\begin{table}
\begin{center}
\caption{Positions and simple estimates of mass from the $\lambda=1.3$ mm continuum. 
\label{sources}}
\begin{tabular}{cccccc}
\tableline\tableline
Source & $\alpha$ (J2000)\tablenotemark{a} & $\delta$ (J2000)\tablenotemark{a} & Flux (Jy) & Mass (\msun) & \=n$_{H_2}$\tablenotemark{b} (cm$^{-3}$) \\
\tableline
L1448 IRS 3A...... & 03 25 36.532 & +30 45 21.35 & 0.196$\pm$0.019 & 0.21$\pm$0.02 & 3.7$\times$10$^7$ \\
L1448 IRS 3B...... & 03 25 36.339 & +30 45 14.94 & 1.094$\pm$0.027 & 1.15$\pm$0.03 & 4.8$\times$10$^7$ \\
L1448 IRS 3C...... & 03 25 35.653 & +30 45 34.20 & $<$ 0.031 & $<$ 0.03 & $<$ 8.1$\times$10$^7$ \\
\tableline
\end{tabular}
\tablenotetext{a}{The positions are from \citet{looney2000}.}
\tablenotetext{b}{Mean number density represented by hydrogen molecules. 
The volumes are estimated as spheres with diameters of $5\arcsec$, $8\arcsec$, and $2\arcsec$
for 3A, 3B, and 3C, respectively.}
\end{center}
\end{table}

\clearpage
\begin{table}
\begin{center}
\caption{Mass, momentum, and kinetic energy estimates of outflow lobes. \label{estm_outflow}}
\begin{footnotesize}
\begin{tabular}{lccccc}
\tableline\tableline
Lobes\tablenotemark{a} & Red-North & Red-East & Black-North & Black-South\tablenotemark{b} & Blue-West \\
\tableline
Outflow Sources & IRS 3A    & IRS 3B   & IRS 3A & IRS 3A \& 3B & IRS 3B \\
\tableline
Integrated Flux (Jy \kms) & 47.47$\pm$1.43 & 101.6$\pm$1.55 & 168.7$\pm$1.75 & 347.3$\pm$2.01 & 304.2$\pm$4.36 \\
Mass ($10^{-3} \textrm{ \msun}$) & 0.153$\pm$0.005 & 0.327$\pm$0.005 & 0.543$\pm$0.006 & 1.12$\pm$0.006 & 0.980$\pm$0.014 \\
Mean no. density\tablenotemark{c} ($10^3$ cm$^{-3}$) & 6.6 & 6.9 & 7.0 & 5.9 & 9.2 \\
Momentum\tablenotemark{d} (10$^{-3} \textrm{ \msun~\kms}$) & 1.2 & 3.5 & 0.43, $-$0.64\tablenotemark{f} & 1.1, $-$1.1 & $-$18 \\
Kinetic Energy\tablenotemark{e} (10$^{41}$ erg) & 1.0 & 4.4 & 0.22 & 0.44 & 38 \\
\tableline
\end{tabular}
\end{footnotesize}
\tablenotetext{a}{Components in Figure \ref{com0map} as contour colors and positions.}
\tablenotetext{b}{This lobe has two components, one from the IRS 3A outflow and the other from the IRS 3B outflow.}
\tablenotetext{c}{Mean number density represented by hydrogen molecules. 
Cylinders along outflows are assumed to estimate the volumes. 
The assumed diameters and lengths of the cylinders are $5\farcs5\textrm{ \& }13\farcs5$, 
$6\farcs0\textrm{ \& }19\farcs0$, $8\farcs0\textrm{ \& }17\farcs5$, $11\farcs0\textrm{ \& }23\farcs0$, 
and $10\farcs0\textrm{ \& }15\farcs5$.
Note that these are not mean number densities indicating 
the whole outflow lobes, but partial components of the lobes. For example,
the mean number density of the northern lobe of the IRS 3A outflow would be 
$6.6+7.0=13.6$ ($10^3$ cm$^{-3}$), assuming the two components occupy the same region.  }
\tablenotetext{d}{Inclination factor, 1/cos$\theta$, is not applied.}
\tablenotetext{e}{Inclination factor, 1/cos$^2\theta$, is not applied.}
\tablenotetext{f}{Plus value is estimated from the redshifted channel and minus from the blueshifted channel.}
\end{center}
\end{table}

\clearpage

\clearpage

\clearpage

\clearpage
\begin{figure}
\includegraphics[angle=270, width=0.95\textwidth]{f4.eps}
\caption{CO $J=2\rightarrow1$\ channel maps of L1448 IRS 3. Two lines indicate the outflow directions that 
are discussed in \S \ref{sec_outflows}. Source locations and beam size are marked.
The synthesized beam is $4\farcs5\times2\farcs5$ and $\textrm{PA}=-2.4 \arcdeg$.
Contour levels are 2.8, 4, 5.7, 8, 11.3, 16, 22.6, 32, and 45.3
times $\sigma=0.144 \textrm{ Jy beam}^{-1}$. \label{cochmap}}
\end{figure}

\clearpage
\begin{figure}
\includegraphics[width=0.71\textwidth, angle=270]{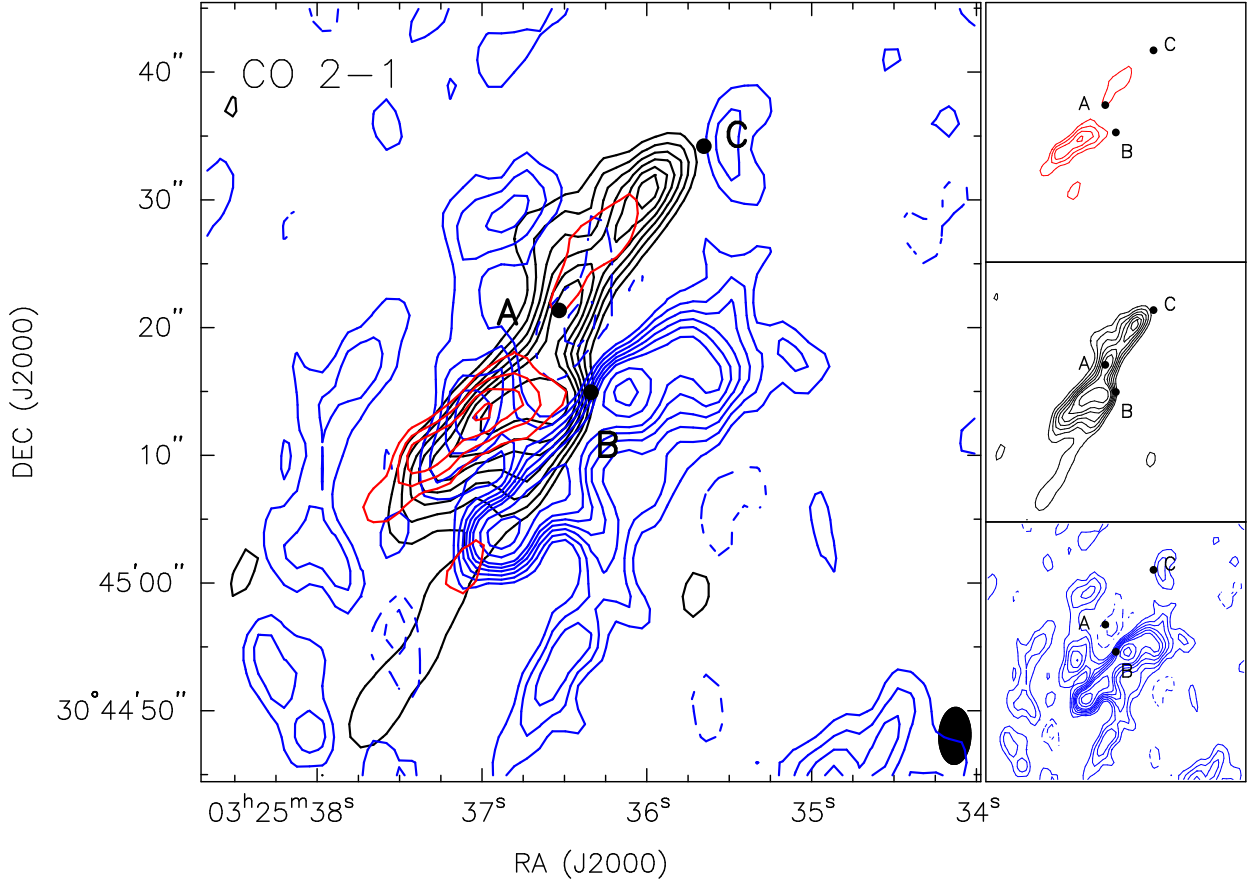}
\caption{Integrated intensity map of L1448 IRS 3. Red, black, and blue
contours present velocity ranges from $+25$ to $+9 \textrm{ \kms}$ (4 channels),
from $+9$ to $+1 \textrm{ \kms}$ (2 channels), and from $+1$ to $-32 \textrm{ \kms}$ (8 channels),
respectively.
The three sub-images on the right have the same velocity ranges
as the main panel with the same contour levels, size-scale, etc., but they are
separated for easier comparison.
The synthesized beam is $4\farcs5\times2\farcs5$~and $\textrm{PA}=-2.4 \arcdeg$.
Black contours mainly show the IRS 3A outflow and red and blue contours mainly represent
redshifted and blueshifted lobes of the IRS 3B outflow.
Blue contours look complicated due to blueshifted components of the mm source
outflow.
Contour levels are 3, 5, 7, 9, 11, 13, 17, 21, 25, 29, 35, 41, and 49 times
$2.3 \textrm{ Jy beam}^{-1} \textrm{ \kms}$. \label{com0map}}
\end{figure}

\clearpage
\begin{figure}
\begin{center}
\includegraphics[angle=270, width=0.75\textwidth]{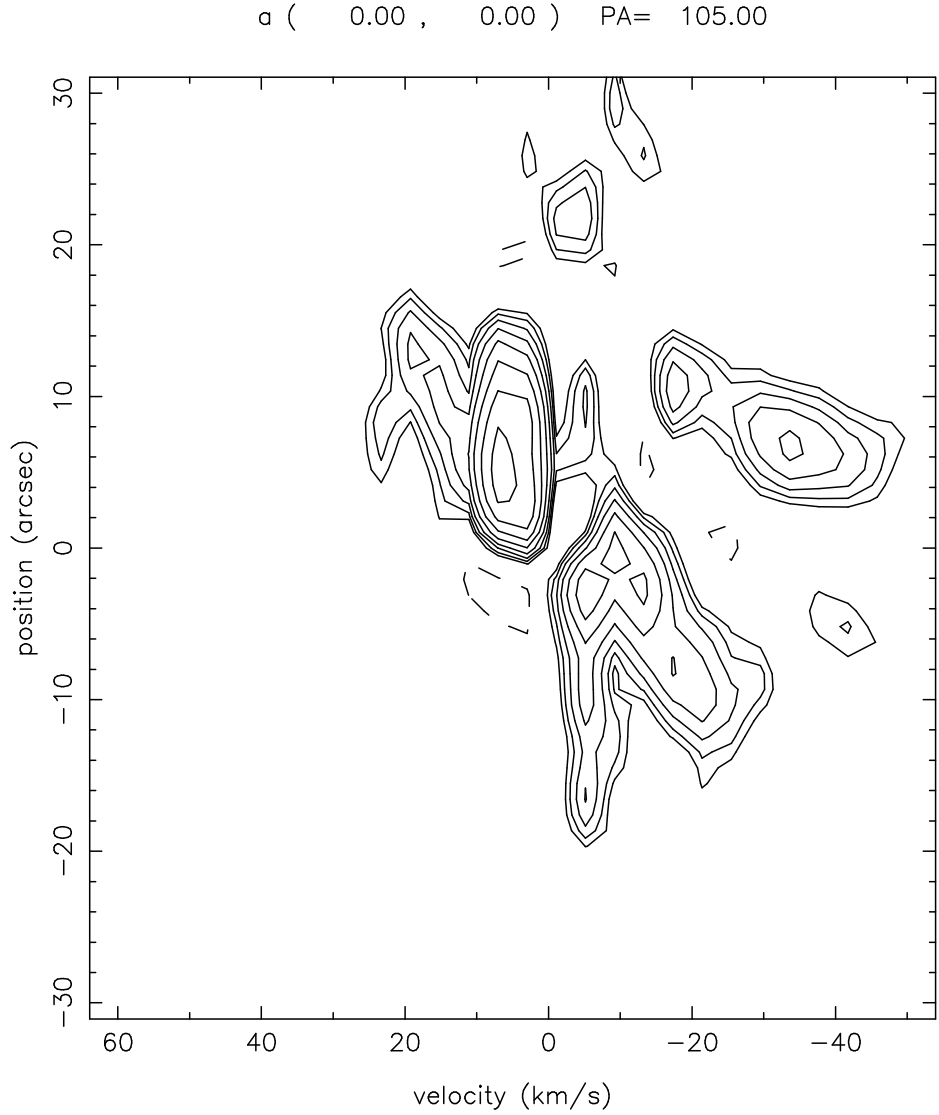}
\end{center}
\caption{Velocity-position diagram of the L1448 IRS 3B outflow.  The cut is along
$105 \arcdeg$ from IRS 3B. Contour levels are 2.8, 4, 5.7, 8, 11.3, 16, 22.6, 32, and 45.3
times $\sigma=0.144 \textrm{ Jy beam}^{-1}$ same as Figure \ref{cochmap}.
\label{vpdiag}}
\end{figure}

\clearpage
\begin{figure}
\includegraphics[width=1.0\textwidth]{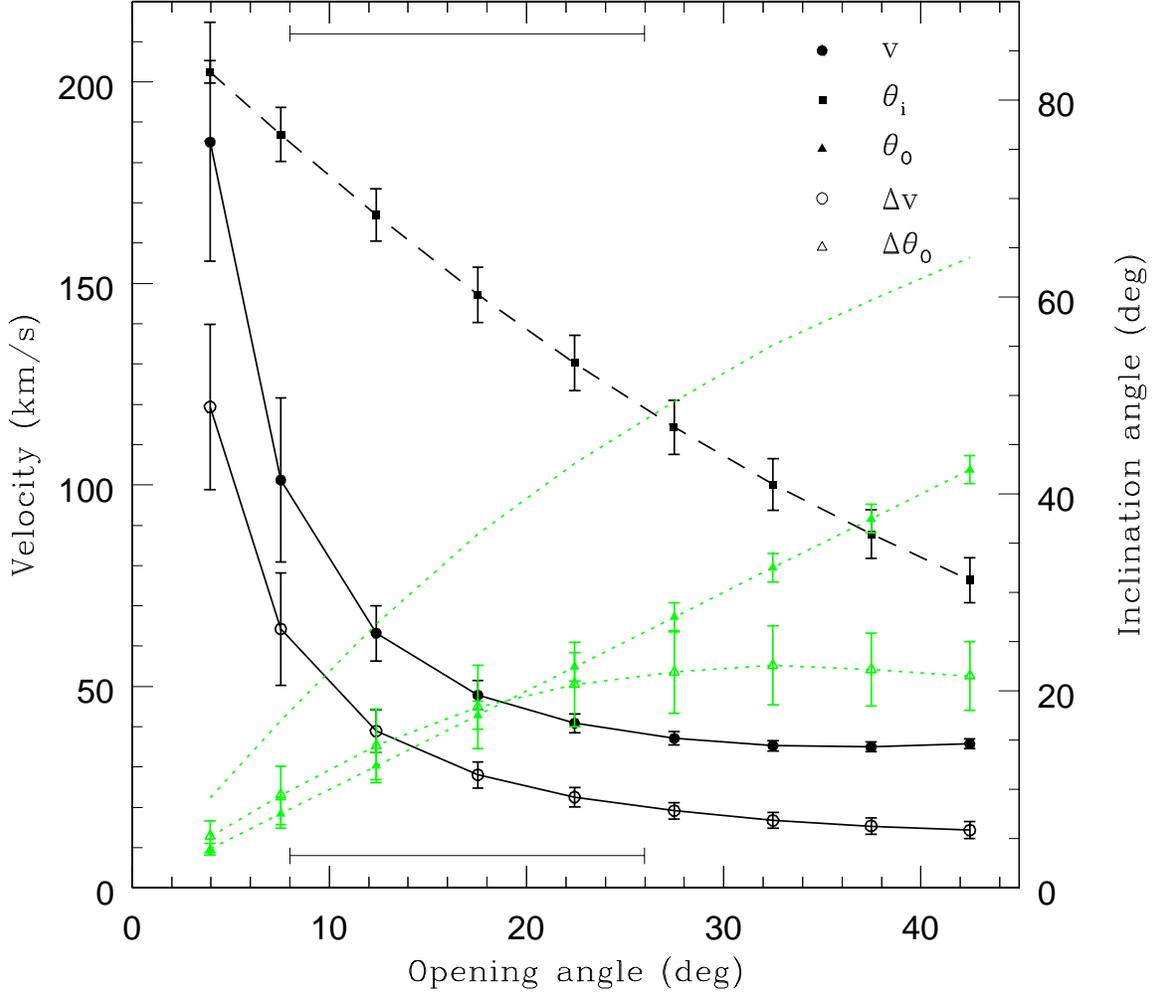}
\caption{
Results of searching IRS 3B outflow parameters, velocity (solid circles), 
inclination angle (solid squares), 
and velocity and opening angle differences (open circles and open triangles, respectively)
versus opening angle.
The opening angle is also plotted (solid triangles) to compare with the other parameters.
The data points are average values of parameters of samples in $5 \arcdeg$ bins of
the opening angle and the error bars present the standard deviations of the bins.
The dotted line without data points indicates the opening angle plus the opening angle difference.
The range of the derived opening angles that are consistent with the observations
($8 \arcdeg < \theta_o < 26 \arcdeg$) is indicated by the horizontal bars at the top and
bottom of the plot.
\label{fig_para}}
\end{figure}

\clearpage
\begin{figure}
\includegraphics[width=1.0\textwidth]{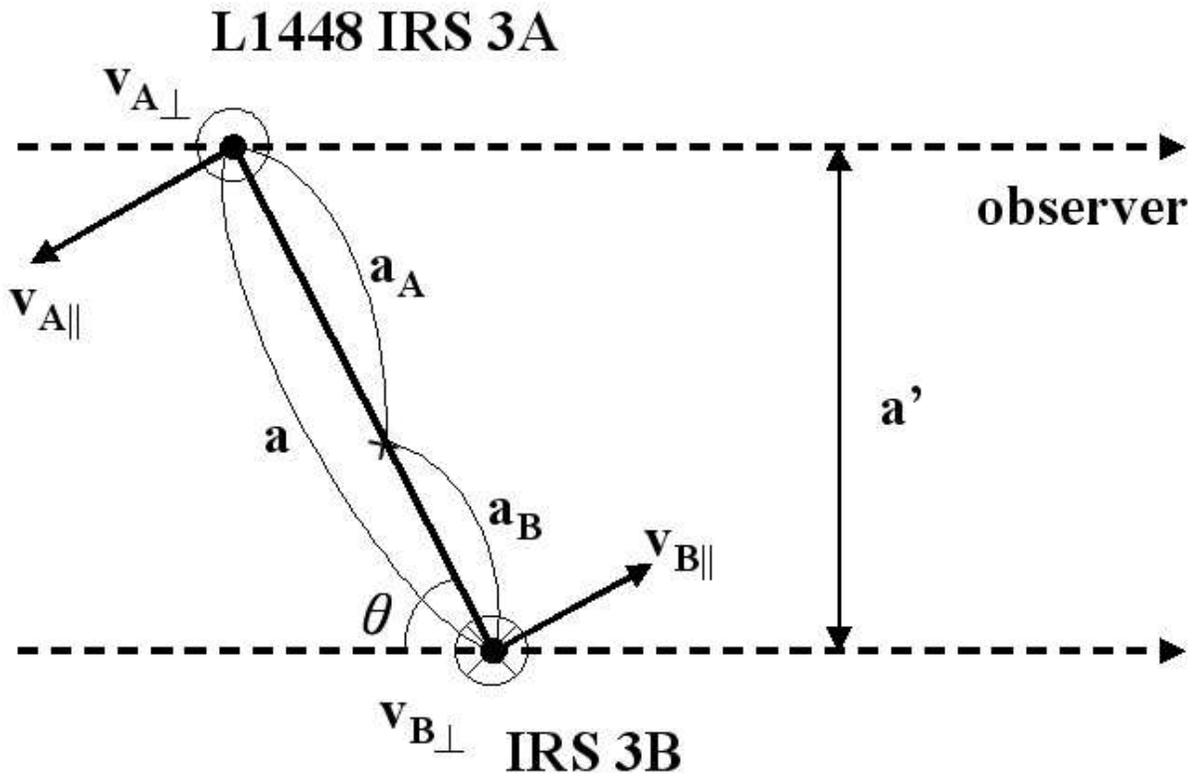}
\caption{Schematic diagram illustrating the binary system of L1448 IRS 3A and 3B.
The velocity components vertical to the line-of-sight plane are assumed 
as forward $v_{A\perp}$ and backward $v_{B\perp}$. They may be opposite directions
such as backward $v_{A\perp}$ and forward $v_{B\perp}$.
\label{fig_sbinary}}
\end{figure}

\clearpage
\begin{figure}
\includegraphics[width=0.7\textwidth, angle=270]{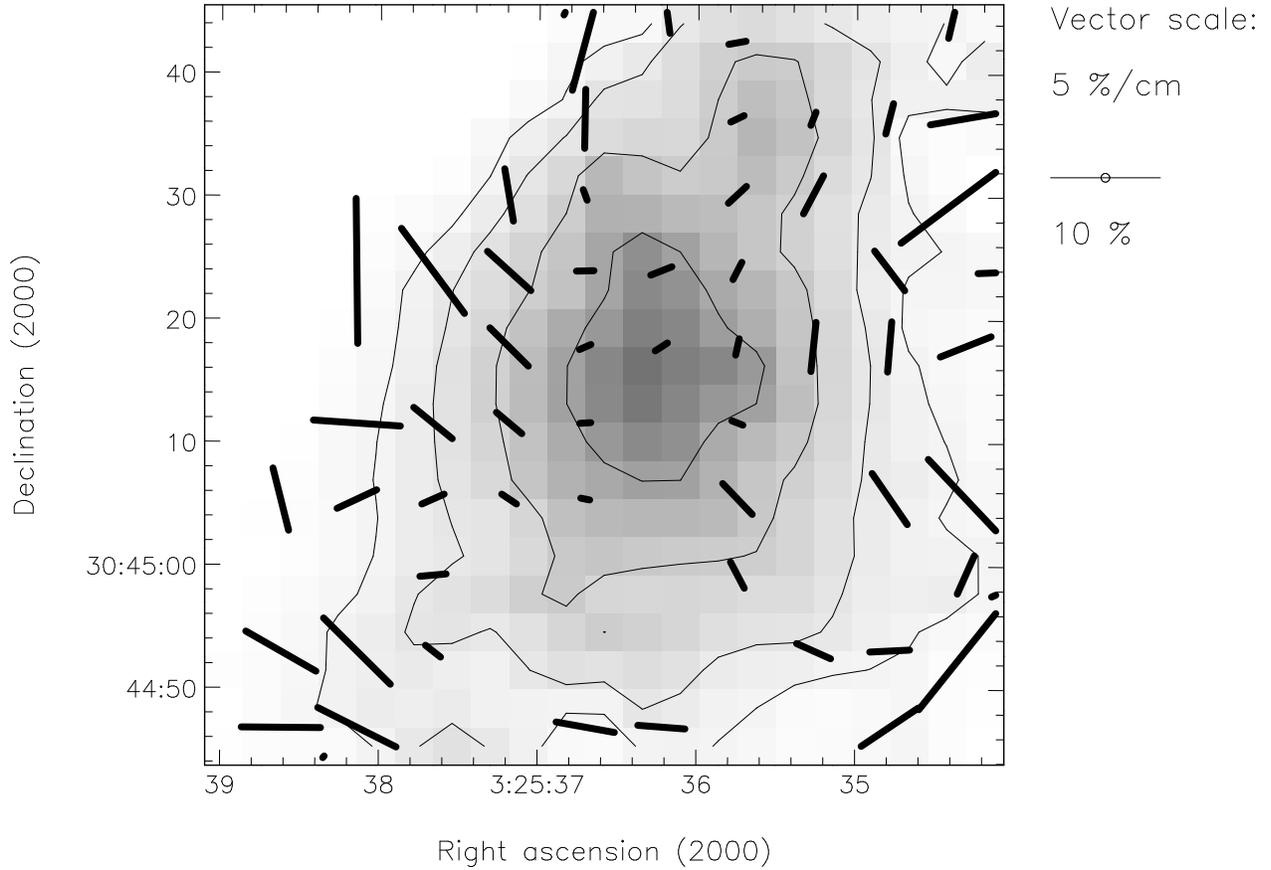}
\caption{Large scale magnetic field in L1448 IRS 3 observed by SCUBA at
$\lambda=850~\mu$m. Note that vectors indicate linear polarization and
the direction around IRS 3B is consistent with our $\lambda=1.3$ mm continuum data.
The beam size is $\sim 13''$.
Gray scales and contour levels are 0.9, 0.8, 0.6, and 0.4 of
the peak intensity, $6.5 \textrm{ Jy beam}^{-1}$ derived from the data presented in
\citet{hatchell2005}. 
\label{scubamap}}
\end{figure}

\clearpage
\begin{figure}
\includegraphics[width=0.8\textwidth, angle=270]{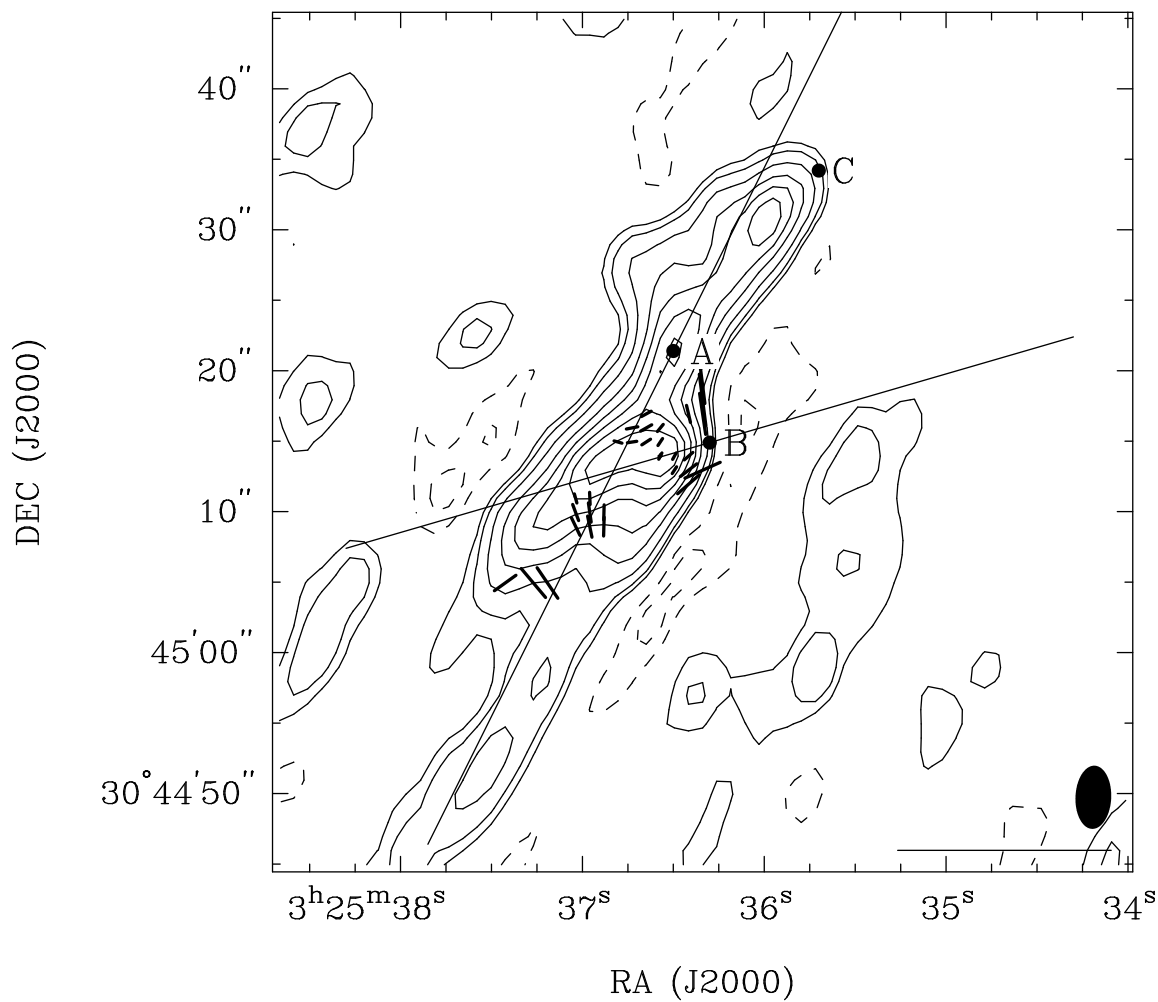}
\caption{CO $J=2\rightarrow1$\ map of L1448 IRS 3, combined in two channels, a velocity range from
 $+1$ to $+9 \textrm{ \kms}$.
Vectors indicate linear polarization and the symbols at right bottom show the
synthesized beam ($4\farcs5\times2\farcs5$~and $\textrm{PA}=-2.4 \arcdeg$) and 100\% polarization scale. 
Two lines present outflow directions
from IRS 3A and 3B. Contour levels are 3, 5, 9, 15, 21, 27, 33, 39, and 45
times $\sigma=0.126 \textrm{ Jy beam}^{-1}$. \label{copolmap}}
\end{figure}

\end{document}